\documentclass[twocolumn,twoside,slac_two]{revtex4}
\usepackage{graphicx}
\usepackage{fancyhdr}

\usepackage{url}
\usepackage{breakurl}
\usepackage[breaklinks]{hyperref}
\hypersetup{
    colorlinks=true,
    urlcolor=magenta,
}

\usepackage{comment}

\begin{document}

\title{Experimental overview on Future Solar and Heliospheric research}

%

\author{T. Laitinen}
\affiliation{Jeremiah Horrocks Institute, University of Central Lancashire, Preston, UK}

\begin{abstract}
  Solar and heliospheric cosmic rays provide a unique perspective in
  cosmic ray research: we can observe not only the particles, but also
  the properties of the plasmas in which the they are accelerated and
  propagate, using in situ and high-resolution remote sensing
  instruments. The heliospheric cosmic ray observations typically
  require space missions, which face stern competition against
  planetary and astrophysics missions, and it can take up to decades
  from the initial concept proposal until the actual observing of the
  cosmic rays can commence. Therefore it is important to have
  continuity in the cosmic ray mission timeline. In this overview, we
  review the current status and the future outlook in the experimental
  solar and heliospheric research. We find that the current status of
  the available cosmic ray observations is good, but that many of the
  spacecraft are near the end of their feasible mission life. We
  describe the three missions currently being prepared for launch, and
  discuss the future outlook of the solar and heliospheric cosmic ray
  missions.
\end{abstract}

\maketitle

\thispagestyle{fancy}


\section{Introduction}

Cosmic rays originating in our heliospheric environment, and the
galactic cosmic rays affected by the heliospheric plasma, are an
interesting and cosmic ray population to study: We can observe not
only the particles, but often also directly or indirectly the
properties of the particle sources and the turbulent plasmas the particles propagate in. Thus, by studying the solar and
heliospheric cosmic rays, we can improve our knowledge of the
fundamental processed responsible for cosmic ray acceleration and
transport in plasma environments in general. In this way, the
heliosphere is our local laboratory for delving into the the physics
of the cosmic rays. The heliospheric cosmic ray research has also a
practical application: the cosmic rays represent a Space Weather
hazard to humans and technology at Earth and in space. For this
reason, the cosmic ray research has also entered national and
international awareness as significant risk for humans.

As any scientific research topic, the heliospheric cosmic ray research
requires experiments to observe the particles. Due to the Earth's
magnetic field, large part of the energy range of interest for
heliospheric particles precludes the use of instruments at the surface
of the Earth, thus the research requires instrumentation that is
attached to spacecraft. Therefore, the advance of the research area
depends on the availability of current, and the outlook of the future
scientific space missions, and the planned instrumentation.

The purpose of this overview is to outline the current status of the
heliospheric cosmic ray instrumentation. We will first outline in
Section~\ref{sec:solar-heli-cosm} the main research topics in the
heliospheric cosmic rays. We then will present the current status of
the instrumentation in Section~\ref{sec:current-fleet}, and discuss
the future instrumentation in Section~\ref{sec:future-missions}.

\section{Solar and Heliospheric cosmic ray populations}\label{sec:solar-heli-cosm}

The sources of the heliospheric cosmic rays observed in the
heliosphere can be classified into three categories: the solar cosmic
rays, or Solar Energetic Particles (SEPs); interplanetary cosmic rays;
and particles originating from outside our heliosphere. A full review
of these categories is beyond the scope of this overview, we will only
briefly outline these categories below.

The SEPs originate at the Sun and its vicinity, related to the
activity of the Sun, and are observed during and after violent
magnetic eruptions at the Sun. The SEP signatures are seen indirectly
in X-rays and gamma rays at the Sun, and as radio bursts in the corona
and in the interplanetary space~\cite[e.g.][]{Vilmer2005SEPsigns}. The
particles also escape from the Sun, and are observed with in situ
instruments throughout the heliosphere, and, in the high-energy Ground
Level Enhancement (GLE) events, also within the Earth's atmosphere, by
neutron monitors, at the surface of the Earth~\cite{Meyer1956GLE}.

The SEPs have been considered to originate at two acceleration
locations: solar flares, which can have field-aligned electric fields,
strong turbulence and internal shock waves; and coronal and
interplanetary shock waves that have been considered as blast waves
and waves driven by the mass ejected during the eruption, the coronal
mass ejection (CME)~\cite{Reames1999}. A two-class paradigm has been
considered related to the two acceleration schemes, and the
differences in particle abundances and flare and radio signatures in
the so-called impulsive and gradual events \cite{Reames1999}. In the
classification, the impulsive events are rich in heavy ions, and the
acceleration takes place in the narrow flare region, whereas
the proton-rich gradual events are due to wide shock waves driven by
the CMEs. Recently, the two-class paradigm has been challenged with
evidence suggesting that the division to the two classes is not clear
\cite{Cane2010,Cliver2016}. In addition, the very wide SEP events
analysed by multiple spacecraft show that the heavy ions,
traditionally related to impulsive events, can have a wide access to a
wide longitudinal ranges \cite{Wiedenbeck2013,Cohen2014}, which seems
to contradict the idea of acceleration at small flare region. The
questions around the two-class paradigm are currently unsolved, and
subject of continuing research.

Particles are also accelerated in the interplanetary space, further
away from Sun. Some coronal mass ejection driven shock waves are
capable of accelerating particles as they propagate throughout the
heliosphere \cite{Reames1999}. Such CMEs offer a possibility to
observe the acceleration region directly, in situ. Particle
acceleration can also take place in the corotating interaction regions
(CIRs), where fast solar wind stream overtakes a slower solar wind,
which, in the Parker spiral geometry, results in an interaction region
between the two solar wind streams~\cite{Hundhausen1972}. At distances
beyond 1~AU from the Sun, the boundaries of the interaction region
form a pair of shocks, which can accelerate particles to MeV energies
\cite{Reames1999}.

Planetary magnetospheres, and their bow shocks, are also capable of
accelerating charged particles. While the planetary particle research
is outside the scope of this overview, the particles escaping from the
planetary acceleration processes into the interplanetary medium are of
interest to the cosmic ray community. In particular, the Jovian
electrons, accelerated at Jupiter \cite{Simpson1992Jovians}, which
represents a steady-state point source in the heliosphere, have been
used to probe the models of cosmic ray propagation in the heliosphere
\cite[e.g.][]{Ferreira2003}.

Particles are also accelerated at the outer boundaries of the
heliosphere. The anomalous cosmic rays (ACR), first observed in 1973,
are an anomalous component in the spectra of some heavy ion species,
as observed during solar activity minima
\cite{GarciaMunoz1973anomalous}. The ACR have been considered as
pickup ions accelerated at the termination shock
\cite{Pesses1981Anomalous}. However, the recent observations by the
Voyager spacecraft have cast doubt to this view
\cite{Stone2005anomalous}.

Finally, the cosmic rays that originate outside our heliosphere are
also important for the heliospheric cosmic ray research. The galactic
cosmic rays are modulated by the interplanetary turbulence
\cite[e.g.][and references therein]{Potgieter2013modulationreview},
which gives information on the properties of the interplanetary
turbulence, the particle transport within the turbulence, and the
source spectrum of the particles outside the heliosphere. The Voyager
1 is now measuring the interstellar spectrum \cite{VoyagerLISMCR2016},
thus the modulated observations in the inner heliosphere can be used
to improve our understanding of the state of the interplanetary
turbulence, and the physics behind the particle transport in such
turbulent fields.

\section{Current Missions}\label{sec:current-fleet}

In this section, we will investigate our current abilities to measure
the cosmic ray fluxes at different energies to study the different
heliospheric cosmic ray sources. We will start our search from the
surface of the Earth, reaching to outside the heliosphere. The
mission and instrument details are collected in
Table~\ref{tab:currfleet}. 

\begin{table*}
\caption{The current fleet of particle-observing spacecraft.\label{tab:currfleet}}
\begin{ruledtabular}

\begin{tabular}{p{2.5cm}llp{2.5cm}l}
Spacecraft & Launched & Orbit & Instruments & particles \\
\hline
Neutron monitor network\cite{Simpson2000neutronmonitor,Bieber1995spaceshipearth} &  & N/A &  &  \\
Pamela\cite{Pamela2007} & 2006 & Quasi-polar LEO &  & 0.08--20 GeV/n \\
AMS\cite{AMS02} & 2011 & LEO (ISS) &  & 1 GV -- 2 TV \\
GOES, GOES-16\cite{GoesR2016} & various & Geosynchronous & SEISS & e:$>$0.6 MeV, p, He: 1-700 MeV/n \\
WIND & 1994 & L1 (current) & 3DP\cite{WIND3DP1995} & e: 3 eV -- 400 keV \\
SOHO & 1995 & L1 & ERNE\cite{ERNE1995}, COSTEP\cite{COSTEP1995} & e: 0.25--10 MeV, ions: 1-120 MeV/n \\
ACE\cite{ACE1998} & 1997 & L1 & CRIS, ULEIS, SWIMS, EPAM, SIS, SEPICA, SWEPAM, SWICS & e: 0.03--0.31 MeV, ions: 0.1 keV--500 MeV \\
STEREO A/B & 2006 & $~$ 1AU heliocentric & IMPACT\cite{StereoImpact2008}(HET, LET, SEPT, SIT) & e: 0.03--6 MeV, ions: 0.06--100 MeV/n \\
Voyager 1/2 & 1977 & 137 AU/113 AU & LECP\cite{VoyagerLECP1977}, CRS\cite{VoyagerCRS1977} & e: 4 eV - 100 MeV, ions: 10 eV - 550 MeV/n \\
\end{tabular}

\end{ruledtabular}
\end{table*}

Cosmic rays have been observed at the Earth's surface using neutron
monitors since 1948 \cite[e.g.][]{Simpson2000neutronmonitor}. Rather
than measuring the incident cosmic ray, these monitors detect the
shower of secondary particles, and their analysis depends on the
atmospheric conditions. In addition, the geomagnetic field affects the
accessibility of the primary particles, with the rigidity cutoff
1-20~GV depending on latitude. Neutron monitors have been widely used
to observe galactic cosmic rays, modulation and solar energetic
particles \cite[e.g.][]{Meyer1956GLE,Simpson2000neutronmonitor,Potgieter2013modulationreview}

Spaceborne cosmic ray instruments can be used to avoid the
complications in observation analysis due to Earth's atmosphere and
magnetic field. However, other complications arise, in particular at
high cosmic ray energies that typically require large instruments,
which are difficult to launch to orbits that allow measurement of both
high and low energy particles. There are currently two spaceborne
instruments that measure the $\sim$~GeV/n energy range. PAMELA
\cite{Pamela2007}, launched to a quasi-polar low-earth orbit (LEO)
with inclination of $70^\circ$, is a large magnetic spectrometer,
which can measure ions at energy range from 80~MeV/n up to several
hundred GeV/n, depending on the orbital phase. AMS \cite{AMS02} is
similarly at LEO orbit, but located at the International Space Station
at lower inclination, 52$^\circ$. Thus, its rigidity cutoff is larger,
500~GV, with the range continuing up to 2~TV.

Above the LEO orbits, the effect of geomagnetic cutoff is reduced as
the spacecraft leave Earth's magnetic field. On the other hand, at
higher orbits, the ability of measuring high-energy particles is
reduced, due to the weight limitations for instruments as the missions
get farther away from the Earth. At geosynchronous orbit the GOES
satellite series has been upgraded with the new GOES-R satellite,
which was recently renamed as GOES-16. GOES-16 provides ion
measurements at considerably lower energies than the LEO instruments,
with the SEISS instrument providing measurements of protons in the
range of 30 eV to 700~MeV, with electrons up to 12 MeV, and heavy
elements up to 200 MeV \cite{GoesR2016}. The spacecraft remains mostly
within the magnetopause, however ions down to ~4 MeV/n have been shown
to be able to penetrate the geosynchronous orbit
\cite{Lanzerotti1968}, depending on orbital and geomagnetic
conditions. Thus, the proton observations of 1--700~MeV are typically
used in cosmic ray analysis, while the lower energy instruments are
designed for magnetospheric particle analysis.

The effect of the Earth's magnetic field on the cosmic ray
observations can be avoided by placing the observing spacecraft
further into the interplanetary space. Stable vantage points for
cosmic ray observations are provided by the Lagrange points of the
Earth-Sun system, and L1 has been used for this purpose by several
space science missions. The L1 is used currently by three spacecraft
with particle instruments. The WIND spacecraft, launched in 1995
originally as a geospace mission on a complicated Earth-Lunar-L1 orbit
\cite{WINDPOLAR1995} is now located at L1, and offers electron
measurements in the range of 3~eV--400~keV by the 3DP instrument
\cite{WIND3DP1995}. The ESA and NASA's SOHO spacecraft was launched to
L1 in 1995, with electron and ion measurements at ranges 0.25-10~MeV
and 1-120~MeV/n respectively, with the instruments ERNE~\cite{ERNE1995}
and COSTEP~\cite{COSTEP1995}. Further, in 1997 also NASA's Advanced
Composition Explorer ACE spacecraft~\cite{ACE1998} joined L1, with a
wealth of instruments (see Table~\ref{tab:currfleet}), detecting
electrons between 30--310~keV and ions between 0.1~keV and
500~MeV. All three spacecraft are still in operation, and have
produced valuable research over the two decades of their operation too
numerous to list here.

A wider perspective to cosmic rays in the heliosphere is offered by
the two STEREO spacecraft, launched in 2006, which orbit the Sun at
approximately Earth's distance from the Sun. STEREO-A advances
22$^\circ$ in heliolongitude each year ahead of Earth, while STEREO-B
trails Earth by the same rate \cite{StereomMission2008}. Both
spacecraft have identical set of instruments, including the IMPACT in
situ instrument set, with four particle sensors measuring electrons
between 30~keV and 6~MeV, and ions between 0.06--100~MeV/n. The STEREO
spacecraft, together with instruments at L1, have been extensively
used for SEP event analysis. In particular, the STEREO observations
have given new understanding into the longitudinal extent of the SEP
events, showing that in some events SEPS can have access throughout the
360$^\circ$ of heliolongitude in the inner heliosphere, challenging
our understanding of SEP event physics
\cite[e.g.][]{Dresing2012,Wiedenbeck2013,Cohen2014,Richardson2014}.

The two STEREO spacecraft reached superior conjunction in 2015,
crossing behind the Sun as viewed from Earth. Due to interference from
Sun at this time, communication to the spacecraft was not
possible. After emerging from the conjunction, the contact to STEREO-B
could not be established before August 2016, after which contact was
soon lost again. Recovery operations for STEREO-B are being planned
for later in 2017 \cite{STEREOstatus}.

Finally, we must mention the Voyager~1 and Voyager~2 which, together
with the Pioneer~10 and Pioneer~11 are the most distant man-made
objects. The LECP~\cite{VoyagerLECP1977} and CRS~\cite{VoyagerCRS1977}
instruments onboard the Voyagers measure electrons between 4~eV and
100~MeV and ions between 10~eV and 500~MeV/n. Voyager~1 has reached
the interplanetary space \cite{VoyagerLISMCR2016}, whereas Voyager~2
is in heliosheath, and observed how the cosmic ray intensities change
throughout the heliosphere \cite[e.g.][and many
others]{Stone2005anomalous,VoyagerLISMCR2016}.


\begin{table*}
\caption{Planetary missions with particle instruments capable of solar and heliospheric particle observations.\label{tab:planetary}}
\begin{ruledtabular}

\begin{tabular}{lrlll}
Spacecraft & Launched & Planet & Instruments & particles \\
 &  &  &  &  \\
\hline
Messenger & 2004--2015 & Mercury & EPPS/EPS\cite{MessengerEPPS2007} & e: 20--700 keV, ions 10 keV -- 5 MeV \\
MAVEN & 2013 & Mars & MAVEN SEP\cite{MavenSep2015} & e: 0.025--1 MeV, ions 0.025 -- 12 MeV \\
MSL & 2011 & Mars & RAD\cite{MSLRAD2012} & e: --10 MeV, ions: --100 MeV/n \\
Cassini & 1997 & Saturn & MIMI\cite{CassiniMIMI2004}(LEMMS, INCA) & 0.02--130 MeV \\
Juno & 2011 & Jupiter & JEDI\cite{JunoJedi2013} & e: 0.01--10 MeV, ions: 0.01-100 MeV \\
New Horizons & 2006 & Pluto & PEPSSI\cite{NewHorizonsPEPSSI2008} & e: 25-500 keV, ions: 0.025--1 MeV \\
\end{tabular}

\end{ruledtabular}
\end{table*}

In addition to the above-discussed spacecraft that carry
instrumentation dedicated to cosmic ray research, additional
instrumentation can be found in several planetary missions (Table~\ref{tab:planetary}). The instruments
are typically designed not from the viewpoint of cosmic ray research
but rather from the point of view of planetary and space physics
research. However, some planetary mission instruments have been used
for analysis of SEP events
\cite[e.g.][]{Lario2013,Lario2016MessengerJuno}.

\section{Future missions}\label{sec:future-missions}

The Tables~\ref{tab:currfleet} and \ref{tab:planetary} give a view of
a wide network of spacecraft capable of observing solar and
heliospheric cosmic rays. However, one must note that there are
several issues that threaten to limit the observation capabilities
already in near future. The spacecraft WIND, SOHO and ACE at L1 are
all 20 years old or more, and their missions have been extended
several times. The STEREO B spacecraft is currently not in operation
\cite{STEREOstatus}. Voyager~1 has departed the heliosphere, with
Voyager~1 soon to follow, and the two spacecraft have power only until
2025, with instrument shutdown commencing in 2020
\cite{VoyagerStatus}. Several of the planetary missions are already at
the end of their planned operations, and as the particle instruments
are designed for planetary science, their orbits may not be optimal
for cosmic ray observations. As an exception, The New Horizons
mission, with the PEPSSI particle
instrument~\cite{NewHorizonsPEPSSI2008}, has now passed Pluto and is
continuing to outer heliosphere, where it can continue the Voyagers'
record of observing outer heliosphere cosmic rays.

The most imminent future missions, summarised in
Table~\ref{tab:future}, concentrate on solar research. The ESA's Solar
Orbiter (SoLO) and NASA's Solar Probe Plus (SPP) will be launched in
2018, with the heliocentric orbits taking the spacecraft into the
vicinity of the Sun. SoLO will descend to 0.285--0.91~AU from the Sun,
to a high-inclination orbit up to 34$^\circ$ at the extended phase of
the mission~\cite{SolO2013}. SPP will go much closer to the Sun, with
the closest perihelion at 8.5~R$_\odot$, or 0.04~AU. Both missions
carry in situ instruments, including cosmic ray detectors. The SoLO
particle suite EPD, measuring electrons up to 15~MeV and ions up to
450~MeV/n, is described in detail in \cite{SolOEPD2017}, in this
volume. The SPP's ISIS instrument reaches somewhat lower energies, up
to 5~MeV electrons and 100~MeV/n in ions~\cite{SPPISIS2016}. In
addition to the in situ suite, SoLO will carry a full set of remote
sensing instruments~\cite{SolO2013}. SPP, on the other hand, will only
house a wide-field heliospheric imager in addition to the in situ
instruments~\cite{SPP2016}.

In addition to the two missions entering orbits close to the Sun, also
L1 will receive attention. India will launch its first solar
observatory, Aditya-L1 in 2019--2020, with a full suite of in situ and
remote instruments
instruments. Included in the instrument set is the energetic particle
detector ASPEX~\cite{AdityaAspex2014}, which will measure electrons
between 0.1 and 20~keV and ions between 0.02 and 5 MeV/n.

Several new mission proposals have considered the Sun-Earth system's
L5 point, trailing Earth by 60$^\circ$ on Earth's orbit, as an optimal
point for a solar mission, particularly from the Space Weather
perspective. The L5 mission suggested by Akioka et
al~\cite{Akioka2005L5} concentrated on CME observations with wide field
coronagraph. Subsequent proposals, EASCO \cite{Gopalswamy2011_L5} and
INSTANT~\cite{INSTANT2016}, contained full set of in situ and remote
sensing instruments, including cosmic ray detectors. An L5 mission is
also included in NASA's Heliophysics Science and Technology
Roadmap~\cite{NASARoadmap2014}. However, so far none of the L5 missions
have been selected. Currently, the most advanced concept is the
Carrington-L5 mission~\cite{CarringtonMission2015}, which is an
operational Space Weather mission. It should be noted that as an
operational mission concept, the drive for the mission is on what is
relevant for Space Weather forecasting: such missions do not
necessarily have scientific goals.

Heliospheric boundary and interstellar missions are being
discussed, with mission proposals such as the IHP/HEX proposed for ESA
``Cosmic Vision 2015--2025'' framework
\cite{IHPHEX.2009.EarthMoonPlanets}. In addition, interstellar probe
is included also in the NASA's Heliophysics Science and Technology
Roadmap~\cite{NASARoadmap2014}. No interstellar missions, however, have
been selected as of now.

Also the new planetary missions should be noted. The ESA's Bepi
Colombo mission will carry the SIXS instrument, designed to use X-rays
for planetary analysis, and housing a particle detector to evaluate
the effect of solar energetic particles~\cite{BepiSIXS2010}. While
SIXS's primary task is not in cosmic ray physics, its ability to
observe the 0.1--3~MeV electrons and 1--30~MeV protons may provide an
opportunity for solar energetic particle research.

ESA is also preparing a mission JUICE to Jupiter, planned to be
launched in 2022, and carrying within its Particle Environment package
a Jovian electron detector JoEE, capable of observing electrons
between 25~keV and 1~MeV. It should be noted though that cruise phase
science is not currently planned~\cite{JuiceRedBook2014}, thus the
applicability of the instrument for heliospheric research may be
limited


\section{Summary}\label{sec:summary}

The current fleet of spacecraft observing cosmic rays, summarised in
Table~\ref{tab:currfleet}, is quite extensive, but many of the
spacecraft are at the end of their missions, and new missions are
needed to continue to improve our understanding of solar and
heliospheric cosmic ray physics. There are currently three missions,
SoLO, SPP and Aditya-L1 (Table~\ref{tab:future}, which will provide
continuation for solar and helilospheric cosmic ray research as well
as probing regions of the heliosphere. Aside of these spacecraft, and
the Bepi Colombo's SIXS, there are a few mission concepts being
proposed, and cosmic ray missions are discussed on relevant
Roadmaps. However, the cosmic ray missions are typically competing
against astronomy and planetary missions. Thus, the future of solar
and heliospheric research needs continuing activity from the community
to ensure continuity of solar and heliospheric cosmic ray research
over the coming decades.

\begin{table*}
\caption{Future missions.\label{tab:future}}
\begin{ruledtabular}

\begin{tabular}{llp{2.5cm}ll}
Spacecraft & Launch & Orbit & Instruments & particles \\
 &  &  &  &  \\
\hline
Solar Orbiter & 2018 & Solar,  0.285--0.91 AU, inclination up to $34^{\circ}$ & EPD suite\cite{SolOEPD2017} & e: 2 keV -- 15 MeV, ions: 3 keV/n -- 450 MeV/n \\
Solar Probe Plus & 2018 & Solar, down to 8.5 R$_{\odot}$ & ISIS\cite{SPPISIS2016} & e: 0.02--6 MeV, ions: 0.04--100 MeV/n \\
Aditya-L1 & 2019--2020 & L1 & ASPEX\cite{AdityaAspex2014} & e:0.1--20 keV, ions 0.02--5 MeV/n \\
 &  &  &  &  \\
\end{tabular}

\end{ruledtabular}
\end{table*}

\bigskip 
\begin{acknowledgments}
The Author acknowledges support from the UK Science and Technology
Facilities Council (STFC) (ST/M00760X/1).
\end{acknowledgments}

\bigskip 

\end{document}